\documentclass[fleqn,usenatbib]{mnras}

\usepackage{newtxtext}

\usepackage[T1]{fontenc}
\usepackage[dvipsnames]{xcolor}

\DeclareRobustCommand{\VAN}[3]{#2}
\let\VANthebibliography\thebibliography
\def\thebibliography{\DeclareRobustCommand{\VAN}[3]{##3}\VANthebibliography}

\usepackage{graphicx}	
\usepackage{amsmath}	
\usepackage{amssymb}	
\usepackage{xcolor}
\title[A multiwavelength study of IRAS 09320+6134]{A Comprehensive Multiwavelength Study of the OH Megamaser galaxy IRAS\:09320+6134}

\author[C. Cassanta et al.]{Claudia M. Cassanta,$^{1}$\thanks{E-mail: claudia.cassanta@acad.ufsm.br (CMC)} Rogemar A. Riffel,$^{1}$\thanks{E-mail: rogemar@ufsm.br (RAR)} 
Andrew Robinson,$^{2}$ Preeti Kharb,$^{3}$  \newauthor Thaisa Storchi-Bergmann,$^{4}$ Jack Gallimore,$^{5}$ Dinalva A. Sales,$^{6}$  C. Hekatelyne,$^{7}$ \newauthor  Stefi Baum,$^{8}$  Christopher O'Dea$^{8}$ 
\\
$^{1}$Departamento de F\'isica, CCNE, Universidade Federal de Santa Maria, Av. Roraima 1000, 97105-900, Santa Maria, RS, Brazil\\
$^{2}$ School of Physics and Astronomy, Rochester Institute of Technology, 84 Lomb Memorial Drive, Rochester, NY 14623, USA \\
${^3}$ National Centre for Radio Astrophysics, Tata Institute of Fundamental Research, S. P. Pune University Campus, Post Bag 3,\\ Ganeshkhind, Pune 411 007, India \\
$^{4}$Departamento de Astronomia, Universidade Federal do Rio Grande do Sul, 91501-970, Porto Alegre, RS, Brazil\\
$^{5}$Department of Physics, Bucknell University, Lewisburg, PA 17837, USA\\
$^{6}$ Instituto de Matem\'atica, Estat\'istica e F\'isica, Universidade Federal do Rio Grande, Rio Grande 96203-900, Brazil \\
$^{7}$ Universidad Tecnol\'ogica, Polo Educativo Tecnol\'ogico Rivera, Ruta 5 km 496, CP 40000, Rivera, Uruguay\\
$^{8}$Department of Physics \& Astronomy, University of Manitoba, 30A Sifton Rd., Winnipeg, MB R3T 2N2, Canada\\
}

\date{Accepted XXX. Received YYY; in original form ZZZ}

\pubyear{2025}

\begin{document}
\label{firstpage}
\pagerange{\pageref{firstpage}--\pageref{lastpage}}
\maketitle

\begin{abstract}
We present a multiwavelength study of the gas distribution, kinematics and excitation of the OH megamaser galaxy IRAS\,09320+6134 (UGC\,5101) using Gemini Multi-Object Spectrograph Integral Field Unit, Hubble Space Telescope, and Very Large Array observations. The HST ACS F814W i-band and H$\alpha+[$N\,{\sc ii}$]\lambda\lambda6548,84$ narrow-band images indicate that this galaxy is a late-stage merger. The ionized gas emission in the inner $\sim$2\,kpc radius, traced by the GMOS data, is consistent with two kinematic components: (i) a rotating disk, observed as a narrow component in the emission-line profiles, with velocity dispersion of $\sigma\leq200$\,km\,s$^{-1}$, and (ii) an outflow, traced by a broad component in the emission-line profiles, with $\sigma\geq500$\,km\,s$^{-1}$. The disk component is well reproduced by a model of rotation in a plane with similar orientation to that of the large-scale galaxy disk. The outflow component presents bulk velocities of up to $-500$\,km\,s$^{-1}$ and corresponds to a mass outflow rate of $\dot{M}_o=0.122\pm0.026\text{\,M}_{\odot}\,\text{yr}^{-1}$. Emission-line ratio diagrams indicate that the gas excitation is mainly due to an active galactic nucleus, likely the driver of the outflow. The VLA radio image reveals a dominant radio core with two-sided emission along the NE-SW direction. The radio core’s spectral index and brightness temperature indicate AGN emission, with the extended emission resembling both in morphology and spectral index the emission observed in radio-quiet quasars. Combined with previous similar studies of other OHM galaxies, the present work supports that this phase is linked to the triggering of an AGN, that seems to occur in the final stages of a merger.
\end{abstract}




\section{Introduction}

Understanding the formation and evolution scenarios of galaxies is one of the major concerns in extragalactic astrophysics. Galaxy mergers and interactions are important processes that affect galaxy evolution, as they can lead to the triggering of star formation as well active galactic nuclei in massive galaxies \citep{thaisa2019}. Objects which are typically the result of galaxy mergers are the Luminous ($L_{\text{FIR}}>10^{11}L_{\odot}$) and Ultraluminous ($L_{\text{FIR}}>10^{12}L_{\odot}$) Infrared Galaxies ((U)LIRGs) \citep{sandmira96}. In the context of overall galaxy evolution, the study of (U)LIRGs became relevant after the discovery that they are more abundant than previously thought. In the local Universe, they are as common as quasars, but when we observe at $z > 1$, the number of (U)LIRGs increases, suggesting that these objects played an important role in the formation and evolution of galaxies \citep{darling0708,lonsdale06}.

The turbulent environment of (U)LIRGs, originated by strong interactions and mergers, as pointed out above, can trigger nuclear starbursts (SB), an active galactic nucleus (AGN) or even both these phenomena \citep{sandmira96}. Nuclear starbursts produce significant amounts of infrared radiation when the ultraviolet emission from young, hot stars is absorbed by dust and re-emitted in the infrared, and, in the same way, AGNs produce large amounts of infrared radiation when material begins to rapidly accrete onto their central supermassive black hole \citep{fiorenza14}. Furthermore, as (U)LIRGS are environments rich in molecular gas and dust, especially in their nuclei, starbursts and AGNs can be hidden behind dust, making detection difficult.

Still within the scope of the connection between (U)LIRGs and AGN-starburst activities, an important aspect to consider is the stage of the merger in which the (U)LIRG phase is observed. Early-stage starburst (U)LIRGs, characterized by younger stellar populations, are expected to transition into late-stage luminous AGN (U)LIRGs \citep{netzer07,hou11,ayu23}. A possible evolutionary scenario, for instance, is that major mergers involving gas-rich galaxies lead to the formation of a massive, cool, starburst-dominated (U)LIRG, which is succeeded by a warm phase as a central AGN becomes active within the dust envelope, subsequently heating the surrounding dust. The central AGN then progresses into an optically bright phase as it disperses the surrounding dust \citep{lonsdale06}. Therefore, starbursts appear to dominate early-stage mergers, whereas AGNs appear to be more dominant in late-stage mergers, in addition to their coexistence at some point in the middle of the process. 

(U)LIRGs can also provide the perfect environments for maser emission, microwave amplification by stimulated emission of radiation. Astrophysical masers and megamasers, the latter with luminosities of about 10$^3$\,L$_{\odot}$, can result from the amplification of the radiation emitted by different molecules, such as hydroxyl (OH), water (H$_2$O), methanol (CH$_3$OH) and silicon monoxide (SiO). The hydroxyl megamasers, or simply OH megamasers (OHMs), are a type of luminous extragalactic maser source that emits non-thermal radiation from OH molecules \citep{wu23}. They radiate in the main lines 1665 and 1667\,MHz, in the ground state of OH, with isotropic luminosities between $10-10^4$\,L$_{\odot}$ \citep{roberts21}. Some of the features of OHM include that they are compact sources in high-resolution observations, possibly variable and generally associated with radio continuum emission from arcsecond-scale observations \citep{wu23}. High-resolution observations have revealed that the OH megamaser emission is concentrated within a 100 pc region in the center of galaxies, with the majority of this emission being found in compact regions ranging up to tens of parsecs in size \citep{lo05,hess21}. High infrared luminosity and the presence of high-density molecular gas in the merger nuclei of (U)LIRGs are the main characteristics that favor the emission of OHMs in these environments \citep{huang18}. However, since 2014, the total number of OHMs found in (U)LIRGs has not increased from $\sim120$ \citep{zhang14,roberts24}. Nevertheless, only $\sim20\%$ of (U)LIRGs show OHM activity, with no clear indicators of the conditions that trigger maser activity or distinguish between masing and non-masing (U)LIRGs \citep{roberts24}. This is expected to change soon with the next generation of H\,I surveys, which promise to increase these numbers, as recent studies have shown \citep{morganti06,suess16,roberts21,hess21,glowacki22}.

Many OHM galaxies may present features of AGN and starburst, individually or as a composite system. On one hand, OHMs are frequently linked to starburst events that take place during galaxy mergers, particularly favoring late-stage mergers with closely spaced nuclei, often within a shared envelope \citep{darling0708}. For example, considering an interaction process, the UV radiation produced by starburst events induced by galaxy mergers heats the dense gas molecules, consequently enhancing their emission. Additionally, the infrared radiation emitted by dust, heated by the UV radiation from starburst events, supplies the energy necessary for the formation of OH megamasers through pumping \citep{huang18}. \citet{baan98} performed a spectroscopic study of (U)LIRGs with OHM emission in order to classify their nuclear activity. In a sample of 40 (U)LIRGs, they found that 67.5\% are classified as AGN against 55\% starbursts, considering in these percentages objects with a composite spectrum. It was also observed that AGNs tend to be more prevalent at higher infrared luminosity, consistent with the trend observed in other studies for a general sample of (U)LIRGs \citep{kim98,veilleux99}, which may suggest that the appearance of AGN and the megamaser emission are connected. It is conceivable that the abundance of AGNs in OHMs is greater, given the challenge in their detection, because they are often obscured by dust. 

Here, we present a detailed study of the gas emission structure and kinematics of the inner $\sim2$\,kpc radius of the galaxy IRAS\:09320+6134 (UGC\:5101; hereafter IRAS\:09320). It is part of a sample composed of 70 galaxies with OH megamaser emission for which we are conducting a comprehensive multi-wavelength study that includes data from the Gemini Observatory, images from the Hubble Space Telescope (HST), and observations from the Karl G. Jansky Very Large Array (VLA). The overall goal of the project is to relate the properties of the OH megamaser and the merger and interaction stage of its host galaxy to the presence of starburst and/or AGN activities. At this initial stage, the focus is on a subset of 15 galaxies, and the galaxy IRAS\:09320 is the seventh for which we present our results. In summary, the results obtained so far indicate the presence of AGNs in 5 out of the 6 OHM galaxies studied, along with circumnuclear star-forming regions in most of them.

IRAS\:09320 is a OHM galaxy with estimated luminosity in the infrared of L$_{\text{IR}}=10^{11.95-12.02}$\,L$_{\odot}$ \citep{huang18,oda17,vivian12,hou09}, which puts it on the border between LIRGs and ULIRGs. It has a redshift $\text{z}=0.03937$ \citep{rothberg06}, which corresponds to a distance of $\sim175$\,Mpc, assuming the Hubble constant as $\text{H}_0=67.8$\,km\,s$^{-1}$\,Mpc$^{-1}$. In addition to an OH megamaser \citep{kandalyan05}, this galaxy also hosts a water megamaser, described as one of the most luminous H$_2$O masers ever detected, with $L_{\rm H_2O}>10^2$\,L$_{\odot}$ \citep{zhang06}. Regarding the ionization mechanism of the gas in IRAS\:09320, classifications in the literature include both starburst activity \citep{karachentsev06} and AGN activity \citep{veilleux95,oda17}, or a composite spectrum, implying contributions from both mechanisms \citep{goncalves99,imanishi03,lonsdale03,paredes15,lucatelli24}.

In this paper we delve into a detailed investigation and mapping of the gas distribution, kinematics, and excitation of the inner $\sim2$\,kpc radius of IRAS\:09320, based on integral field spectroscopy (IFS) data from Gemini, broad-band and ramp filter H$\alpha+$[N\,{\sc ii}] images from HST and 1.4\,GHz continuum image from VLA. The paper is organized such that in Section \ref{2}, we present details about the observations and data reduction process; in Section \ref{3}, we provide the overall results of the different observations, focusing on the distribution, kinematics, and ionization of the gas in the galaxy; in Section \ref{4}, we discuss in further detail the kinematics and the nature of the gas ionization mechanism; and finally, in Section \ref{5}, we present our conclusions.

\section{Observations, data reduction and measurements} \label{2}

\subsection{HST data}

The images from HST were acquired through a snapshot survey of a large sample of OH megamaser galaxies (PI: D. J. Axon). Each galaxy in the sample was observed using the Advanced Camera for Surveys (ACS) Wide Field Channel (WFC), employing broad (F814W), narrow (FR656N), and medium (FR914M) band filters. These observational configurations enable qualitative analysis of the observed object. The F814W filter images were obtained for studying the structure of the host galaxy, whereas the ramp filter images facilitate investigation into the distribution of ionized gas. The central wavelengths were chosen to encompass H$\alpha$ in the FR656N filter and the adjacent continuum for subtraction in the FR914M filter. Total integration times were 600 seconds for both the broad-band F814W and the narrow-band H$\alpha$ FR656N filters, and 200 seconds for the medium-band FR914M filter.

The data reduction process was done using the Image Reduction and Analysis Facility ({\sc IRAF}\footnote[1]{{\sc iraf} is distributed by the National Optical Astronomy Observatory, which is operated by the Association of Universities for Research in Astronomy (AURA), Inc., under cooperative agreement with the National Science Foundation.}; \cite{iraf}), as detailed in \cite{sales15,sales19}, and followed the standard procedure for HST images. Cosmic rays were removed using the {\sc lacos} \citep{lacos} task. To obtain a continuum-free H$\alpha+[$N\,{\sc ii}] image of IRAS\,09320, we first calculated the count rates for foreground stars in the medium (FR914M) and narrow (FR656N) band filter images, which allow us to determine a mean scaling factor to be subsequently applied to the medium-band FR914M image and, finally, subtracted from the narrow-band FR656N image. Then, the continuum-subtracted H$\alpha$+[N\:{\sc ii}] image was inspected to verify that there were no residuals at the positions of the foreground stars. Typically, this method shows flux uncertainties with a margin of $\sim5$ to $10$\% (see \cite{hoopes99,rossa00,rossa03}).

\subsection{VLA radio continuum data}

IRAS\,09320 has been observed with the Karl G. Jansky Very Large Array (VLA) on October 9, 2016, as part of program 16B-063 (PI: J. Gallimore). The VLA was in its A-configuration, and the receivers were tuned to L-band frequencies spread over nine (9) intermediate frequency (IF) bands. One IF was reserved for observing redshifted hydroxyl (OH) maser transitions; unfortunately, interference, primarily from global navigation satellites, rendered that IF unusable. The remaining eight (8) IFs observed continuum emission between 1051.5 and 1947.5~MHz (IF central frequencies). Each IF spanned 128~MHz bandwidth in dual polarization (RR and LL). The total observing time was 72.7 minutes with 39.5 minutes devoted to IRAS\,09320. We also observed 3C~147 (13.7 minutes) for flux and bandpass calibration and J0921+6215 (18.5 minutes) for phase calibration.

Primary calibration was performed in CASA 6.5.4-9 using the continuum pipeline provided by NRAO. Imaging and self-calibration were also performed in CASA. For imaging, we used the CASA task {\tt tclean}, employing the {\tt mtmfs} deconvolution algorithm \citep{2011A&A...532A..71R} and Briggs' weighting \citep{1995PhDT.......238B}. The final, restored image is provided in top panel of Fig.~\ref{fig:radio}; the effective frequency of the restored image is 1.563~GHz. To accommodate the $\sim 1\arcsec{}$ synthetic beam and recover background radio sources within the $\sim 0\fdg{}5$ primary beam, we generated $8192 \times 8192$ pixel images with 0\farcs{25} pixels. Self-calibration involved repeated cycles of CLEAN deconvolution down to the $5\sigma$ surface brightness level and phase calibration against the current CLEAN model with descending solution intervals ($\Delta t = $ 10\,min, 30\,s, 10\,s, and 1\,s). After applying the final phase self-calibration, we performed one cycle of amplitude self-calibration with a solution interval $\Delta t = 5$~minutes. The final, restored image achieved a background rms $ = 17\ \mu\mbox{Jy\ beam}^{-1}$ with a $1\farcs{}18 \times 1\farcs{}02$, PA = 4$\degr$ restoring beam.

The {\tt mtmfs} deconvolution algorithm provides a image of the in-band spectral index $\alpha$ ($S_{\nu} \propto \nu^{\alpha}$). The L-band spectral index image is shown in bottom panel of Fig.~\ref{fig:radio}. This image was obtained by blanking the in-band spectral index image using the in-band spectral index ``noise'' image with a threshold of 0.4. We discuss the results ahead in Section \ref{32}.

\subsection{GMOS-IFU observations and data reduction}

Spectroscopic data of IRAS\:09320 were obtained with the Gemini North Telescope. Observations were carried out on the night of January 25, 2022, using the Gemini Multi Object Spectrograph (GMOS) \citep{hook04} operating in the Integral Field Unit (IFU) \citep{allington-smith02} mode. Three observations were performed with the IFU operating in one slit mode using the IFU-R mask and the grating B600-G5307, which has a resolving power of R $ \sim1688$. Each observation had 1200\,s of exposure time, and they were centered on three different wavelengths, at 5900\,\AA, 6000\,{\AA} and 6100\,\AA, in order to avoid important emission lines falling within the gaps between the GMOS CCDs. The field of view with this configuration is $3.5$\,arcsec $\times$ $5.0$\,arcsec and the spectral range covers the strongest optical emission lines, from H$\beta$ to [S\,{\sc ii}]$\lambda6731$\,\AA.

The data reduction process was done with the software \textsc{iraf} together with \textsc{gemini gmos} data reduction packages. It consists of several steps, starting with the bias level subtraction for each image, the removal of scattered light and cosmic rays, the latter using the {\sc lacos iraf} task \citep{lacos}. Next, the images were corrected for the quantum efficiency of the CCD, and flat-fielded, followed by the calibrations in wavelength, using the CuAr lamp spectra, and in flux, using a sensitivity function obtained from the spectrum of a standard star, Feige 66 in our case. The images were also trimmed and sky subtraction was performed.

The final step is the construction of a single data cube, obtained from the median combination of the individual cubes, and using the peak of the continuum emission as reference. The final data cube has a spatial sampling of $0.1\times0.1$\,arcsec$^2$. Also, we estimate the seeing during the observation from the measurement of the full width at the half-maximum (FWHM) of the flux distributions of field stars present in the acquisition image, which is obtained just before the spectroscopic observations, we obtained $\sim0.85$\,arcsec, which corresponds to $\sim0.72$\,kpc in the galaxy, considering a distance of $175$\,Mpc. The velocity resolution, obtained from the measurement of the FWHM of several emission lines of the CuAr lamp spectra, is $\sim90$\,km\,s$^{-1}$.

\subsection{GMOS-IFU data analysis}

\begin{figure*}
\begin{center}
    \includegraphics[width=0.45\textwidth]{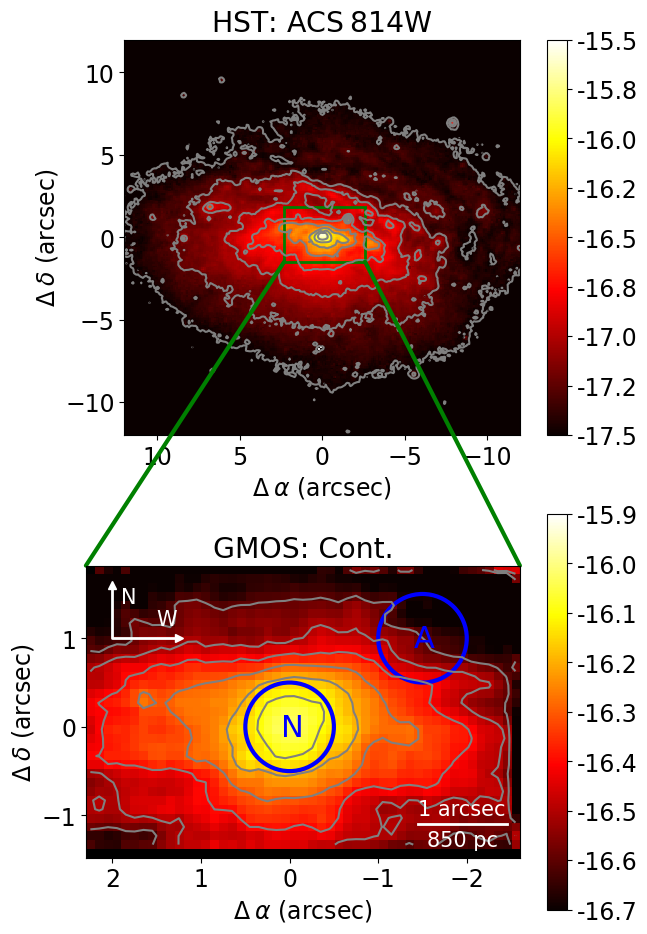}
    \includegraphics[width=0.45\textwidth]{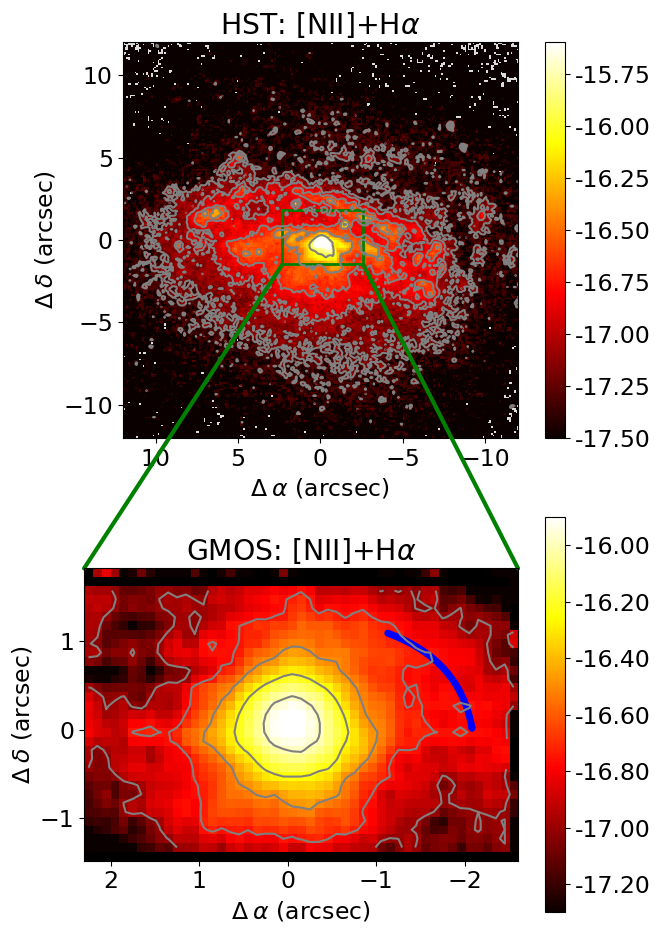}
    \includegraphics[width=0.95\textwidth]{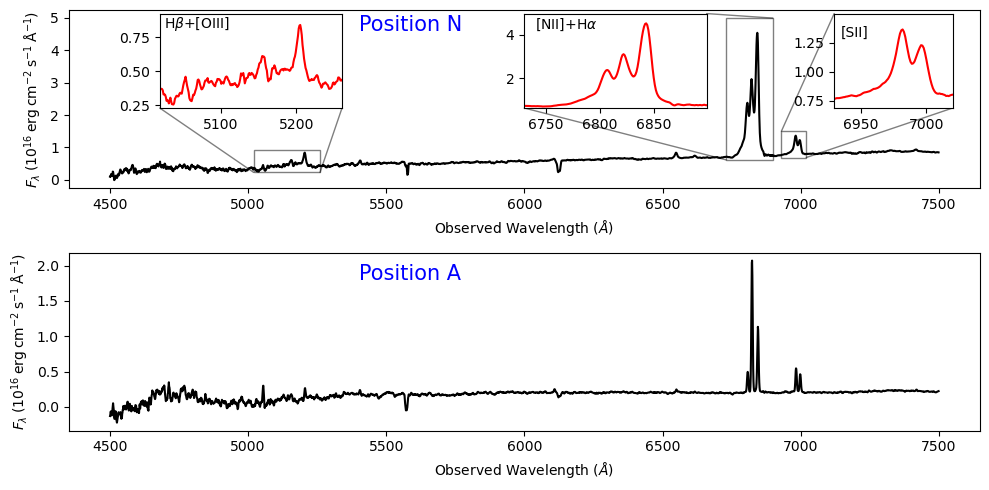}
\end{center}
   \caption{The top panels show the HST ACS F814W - i band image (left) and H$\alpha+[$N\,{\sc ii}$]\lambda\lambda6548,84$ narrow-band image (right) of IRAS\:09320. The central rectangle shows the GMOS field of view. The middle panels show the GMOS-IFU continuum image within a spectral window 5700--6200\,{\AA} (left) and a pseudo narrow-band image covering the H$\alpha+[$N\,{\sc ii}$]\lambda\lambda6548,84$ emission lines (right). The color bars show the continuum in logarithmic units of erg\:s$^{-1}$\:cm$^{-2}$\:\AA\:$^{-1}$\:arcsec$^{-2}$ and the H$\alpha+[$N\,{\sc ii}$]\lambda\lambda6548,84$ fluxes in logarithmic units of erg\:s$^{-1}$\:cm$^{-2}$\: $^{-1}$\:arcsec$^{-1}$. The curved blue line on the GMOS: [N\,{\sc ii}]+H$\alpha$ map outlines the region where we identified a probable spiral arm of the galaxy. In all images, the gray contours represent flux levels of each image.
   The bottom plots display two representative spectra from distinct regions: the nucleus (labeled as N in the GMOS continuum image) an extra-nuclear region (labeled as A), at 1.8\,arcsec northwest of the nucleus. These spectra are integrated within a circular aperture of 0.5\,arcsec radius. The insets in the nuclear spectrum display the regions of the main emission lines.}
    \label{fig:spectra}
\end{figure*}

The bottom panels of Fig.~\ref{fig:spectra} show example spectra of IRAS\:09320 extracted from two different regions of the galaxy: the nucleus (Position N), defined as the position of the peak of the continuum emission, and an extra-nuclear region at 1.8 arcsec northwest from the nucleus (Position A). These spectra were extracted within a circular aperture with radius of 0.5 arcsec.
The main emission lines identified are: [O\,{\sc iii}]$\lambda\,5007$, H$\alpha$, [N\,{\sc ii}]$\lambda\lambda6548,6583$ and [S\,{\sc ii}]$\lambda \lambda6717,6731$. For a more detailed visualization of the emission line profiles in the nuclear spectrum, zooms are shown in the spectral regions of each line individually.

To map the distribution and kinematics of the gas, and subsequently determine the gas ionization mechanism, we fitted the profiles of the emission lines present in the spectra using the \textsc{IFSCube} Python package \citep{ifscube1,ifscube2}. An initial fitting, using Gauss-Hermite series, was performed and, although it reproduced well the observed profiles of the emission lines at some positions in the galaxy, high absolute values were obtained for the Gauss-Hermite moments h$_3$ and h$_4$, indicating the presence of more than one kinematic component. The presence of a blueshifted broad wing in the emission line profiles is clearly seen in the zoom for the nuclear spectrum shown in Fig.~\ref{fig:spectra}. Therefore, we performed the fitting using two Gaussian curves per emission line, one narrow and one broad. We have adopted the usual hypothesis that the narrow component represents the gas emission in the disk, while the broad component represents non-circular motions, associated to outflows in ionized gas \citep[e.g.][]{DallAgnol2021}. During the fitting process, we included the constraint that the velocity dispersion of the broad component must be greater than that of the narrow component and the [N\,{\sc ii}]$\lambda6583/$[N\,{\sc ii}]$\lambda6548$ line ratio was kept fixed at its theoretical value of 3.06 \citep{oster}. The underlying continuum was modeled using a 6th-order polynomial function to approximate it, and this pseudo-continuum was subsequently subtracted to isolate the emission lines.

\begin{figure}
    \includegraphics[width=\linewidth]{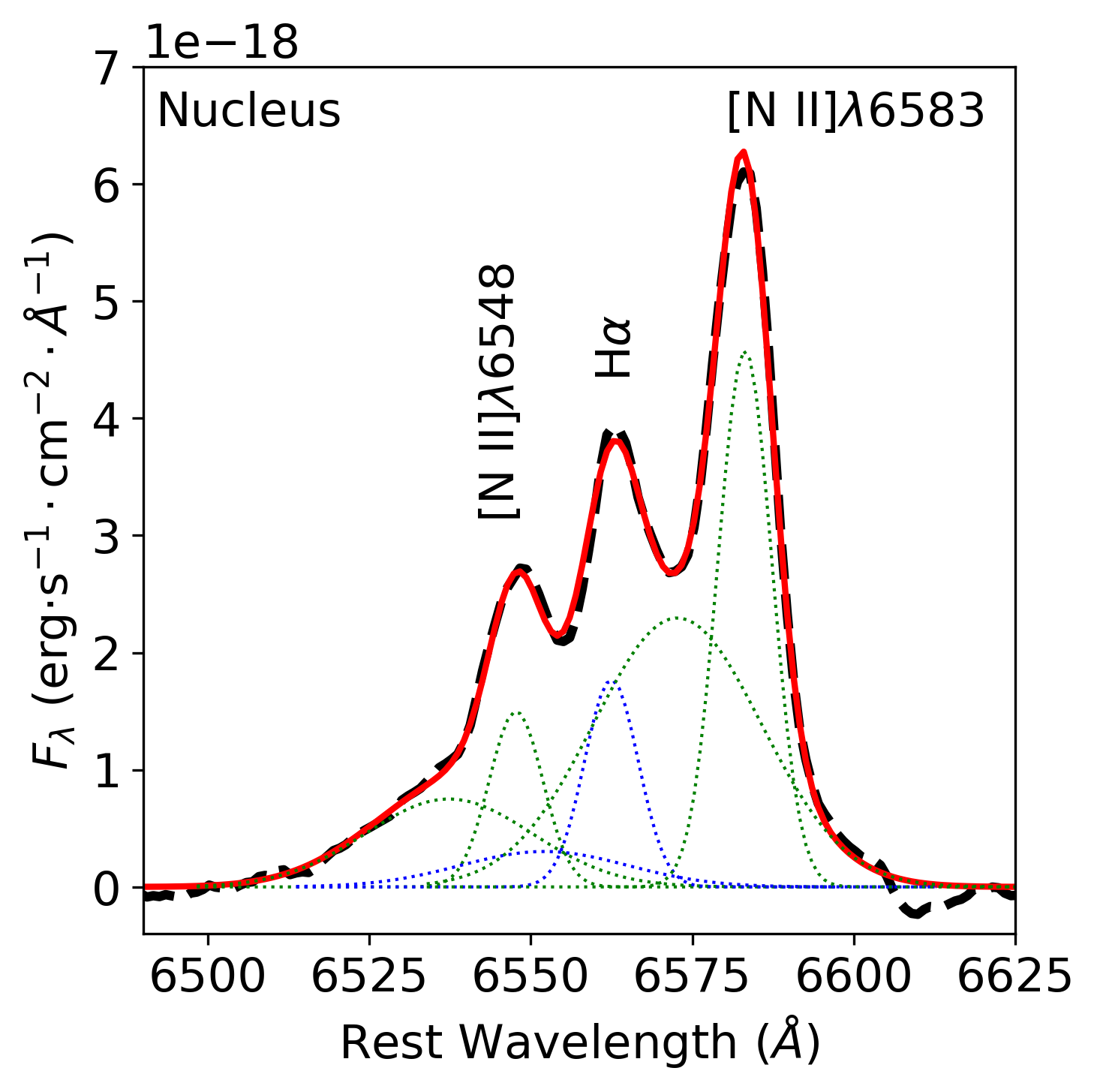}
    \includegraphics[width=\linewidth]{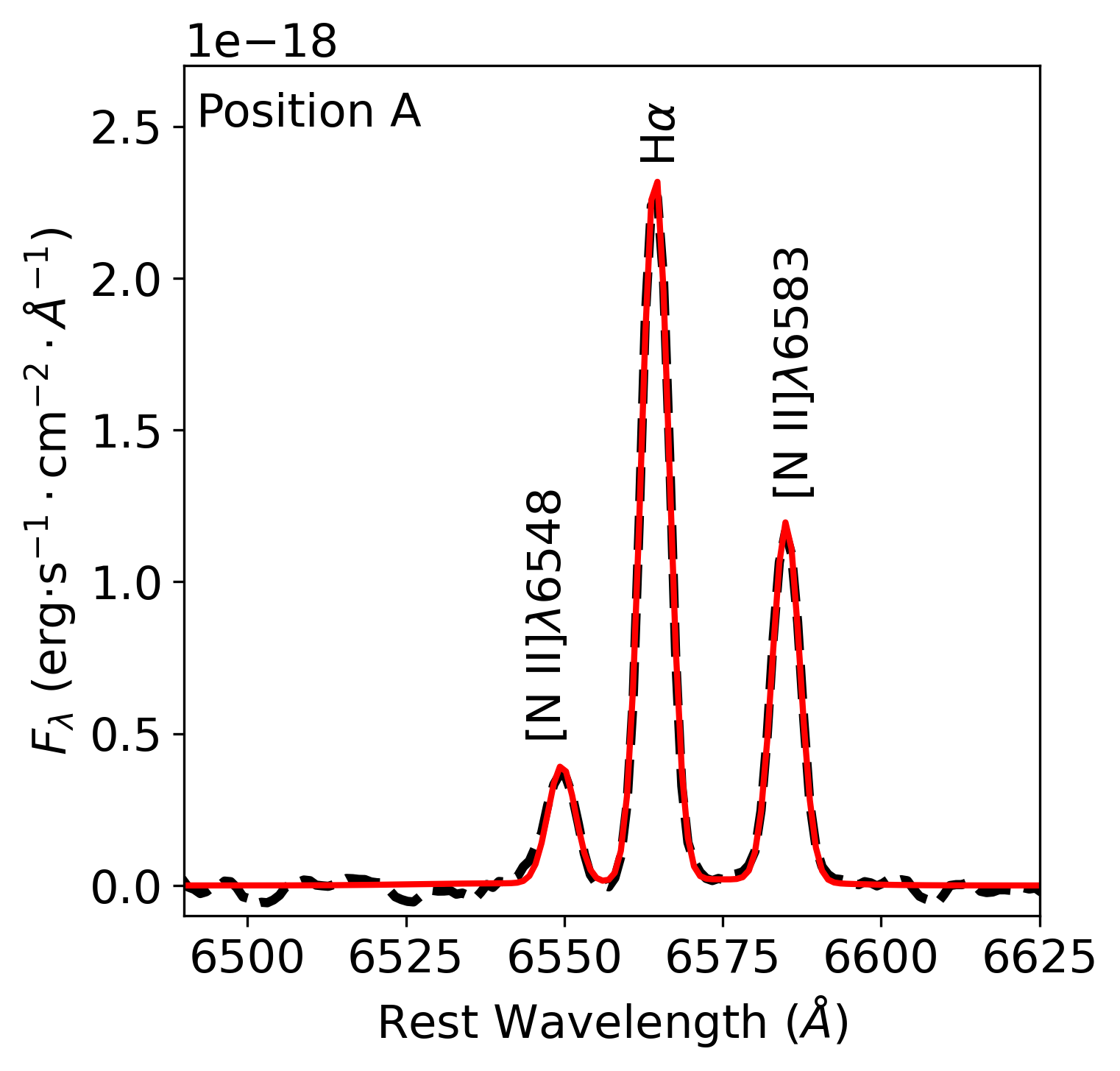}
    \caption{Examples of emission line fitting for the H$\alpha+$[N\,{\sc ii}]$\lambda\lambda6548,6583$ complex. The top panel shows the fit for the nuclear spaxel, with two Gaussian curves per emission line, while the bottom panel displays the fit with a single Gaussian curve per emission line for an extra-nuclear spaxel, located 1.8\,arcsec northwest of the nucleus. The dashed black line represents the observed spectrum and the continuous red line represents the adjusted model. For the nuclear region, the individual Gaussian components are shown as dotted lines, being the blue one for the broad component and the green one for the narrow component.}
    \label{fig:ajustes}
\end{figure}

\section{Results} \label{3}
\subsection{Flux distributions}

The top panels of Figure \ref{fig:spectra} show the ACS/HST F814W i-band and the narrow-band H$\alpha+[$N\,{\sc ii}]$\lambda\lambda6548,6584$ images of IRAS\,09320, respectively. Below, zooming in on the highlighted central region of the HST panels, are the corresponding GMOS-IFU field of view images.

The i-band HST image shows an almost regular structure with an emission peak at the nucleus of the galaxy. This emission structure is more elongated along the east-west direction. The GMOS continuum image, shown in the middle left panel, was obtained by calculating the average fluxes from a spectral region with no strong emission lines. It is consistent with the HST continuum image, also showing a more elongated emission structure from east to west. Considering that only one bright nucleus was detected in this galaxy, it is suggested that IRAS\,09320 is possibly a late-stage merger. This classification agrees with an ongoing work (Wills et al., in prep.) on the same object using long-slit data, where IRAS\,09320 is classified as a post-merger due to the single bright nucleus and the presence of a low surface brightness morphology on large scales ($\sim20$\,kpc).

The HST continuum-free H$\alpha+[$N\,{\sc ii}]$\lambda\lambda6548,6584$ narrow-band image, presented in the top right panel of Fig. \ref{fig:spectra}, shows a similar flux distribution to that of the i-band continuum image, with emission peaking at the galaxy's nucleus, but here we can also see some knots of emission surrounding the nucleus at distances of 2--7\,arcsec. The GMOS pseudo-narrow-band image, obtained by integrating the fluxes in the H$\alpha+[$N\,{\sc ii}]$\lambda\lambda6548,6584$ region and subtracting the continuum from adjacent regions, is shown in the middle right panel. It shows extended emission over the entire field. It also shows enhanced emission in regions surrounding the galaxy's nucleus, at distances of $\sim2$\,arcsec from it, which seem to delineate partially spiral arms to the northwest and southeast. Analyzing both the HST and GMOS images together with the spectra shown in the lower panels, this extended emission structure is most likely associated to circumnuclear star-forming regions, as evidenced by the strong H$\alpha$ emission relative to the [N\:{\sc ii}] emission lines. 

Fig. \ref{fig:ajustes} presents examples of emission line fitting for the H$\alpha+$[N\,{\sc ii}]$\lambda\lambda6548,6583$ complex. The top panel shows the fit for the nuclear spaxel (Position N in Fig. \ref{fig:spectra}), while the bottom panel displays the fit for a spaxel in an extra-nuclear region, located 1.8\,arcsec northwest of the nucleus (Position A in Fig. \ref{fig:spectra}). In the nuclear spectrum, both broad and narrow Gaussian components can be identified fitting the emission line profiles, whereas in the extra-nuclear region, the profiles are well modeled by a single Gaussian component.

\begin{figure*}
    \includegraphics[width=\textwidth]{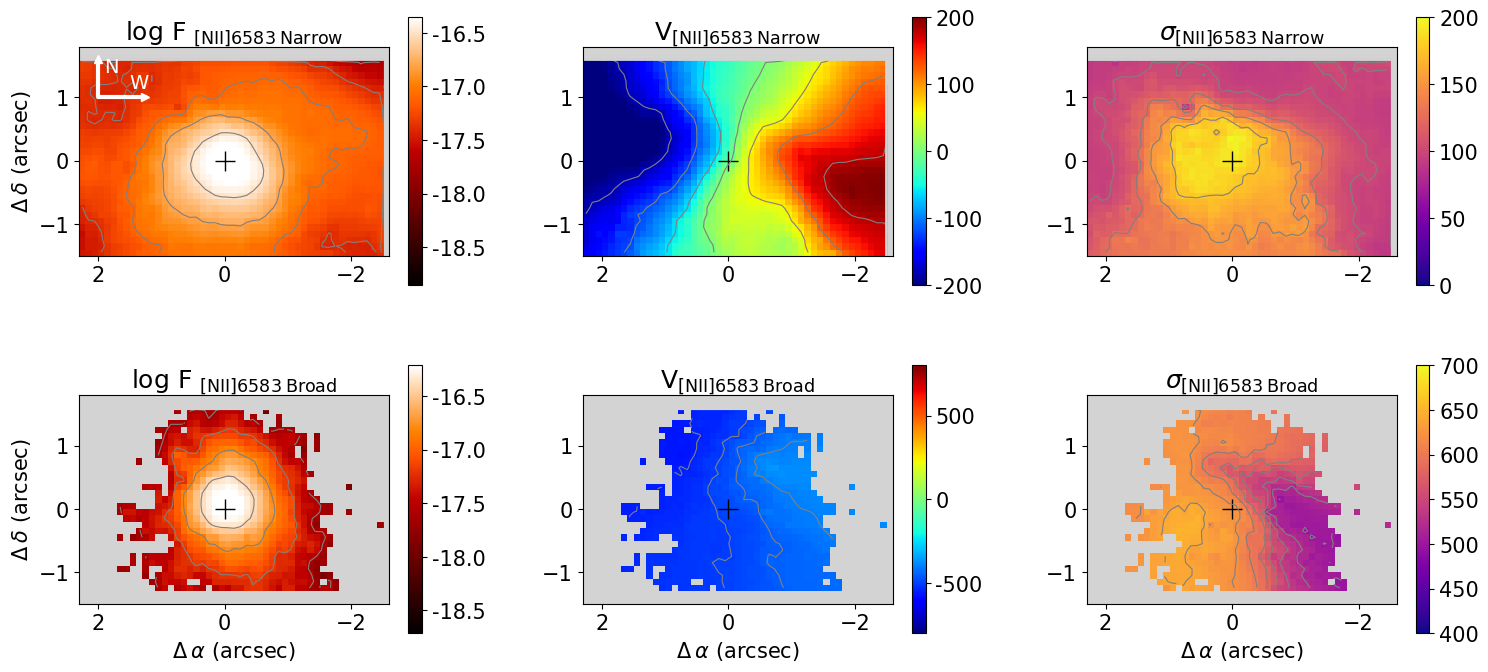}
    \caption{Maps of the distribution and kinematics of the gas in the galaxy IRAS\:09320, for the emission line [N\,{\sc ii}]$\lambda6583$. From top to bottom, we have the maps for the narrow component and for the broad component. Flux distributions are shown in the first column, in logarithmic units of erg\,s$^{-1}$\,cm$^{-2}$\,spaxel$^{-1}$. In the second and third columns are the maps of velocity fields and velocity dispersion, respectively, in units of km\,s$^{-1}$. Velocity fields are shown after subtraction of the galaxy's systemic velocity. In all maps, the central cross marks the position of the galaxy's nucleus, and gray areas correspond to masked regions where the emission line was not detected or the signal-to-noise ratio was not high enough to allow measurements.}
    \label{fig:nii}
\end{figure*}

A more detailed analysis of the emission-line flux distributions in the central region of IRAS\,09320 can be obtained from the flux maps of individual lines, obtained from the GMOS datacube. The first column of Fig.~\ref{fig:nii} presents the flux distributions for the narrow (top panel) and broad (bottom panel) components of the [N\,{\sc ii}]$\lambda6583$ emission line. We chose this specific line because it is the strongest one observed in the nuclear region of of IRAS\:09320. In all maps, the fluxes are shown in logarithmic units of erg\,s$^{-1}$\,cm$^{-2}$\,spaxel$^{-1}$ and the gray regions correspond to masked locations where the emission lines amplitudes are lower than $3\sigma$ of the adjacent continuum.

Both the narrow and broad components exhibit their emission peaks at the nucleus, defined as the position of the continuum peak. The narrow component emission for all lines is observed across the entire GMOS field of view, except for [O\:{\sc iii}]$\lambda$5007, which shows some spaxels where its emission is not detected at distances greater than 1\,arcsec ($\sim0.85$\,kpc) from the nucleus. In addition, there is a structure resembling a spiral arm to the northwest from the nucleus, clearly observed in the H$\alpha$ flux map and already seen in the HST maps (Fig. \ref{fig:spectra}), attributed to the emission from star-forming regions.

The broad component emission is restricted to the inner 2\,arcsec ($\sim1.7$\,kpc) radius for the [N\:{\sc ii}]$\lambda$6583 emission line, and less extended in other emission features. Its flux distribution morphology is approximately round and centrally peaked, slightly more elongated towards the southwest. The [O\:{\sc iii}]$\lambda$5007 broad component is detected only in a few spaxels close to the nucleus of the galaxy.

\begin{figure}
\centering
 \includegraphics[width=0.9\linewidth]{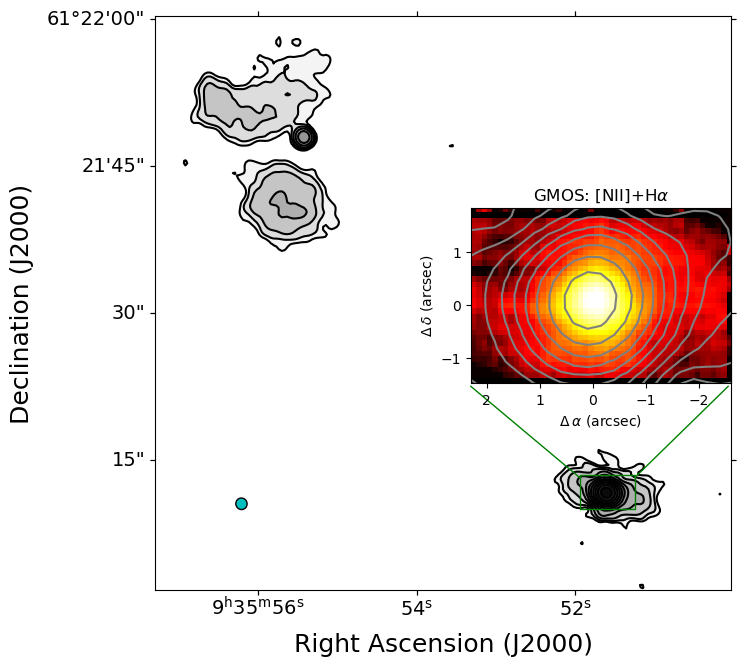}
    \includegraphics[width=0.91\linewidth]{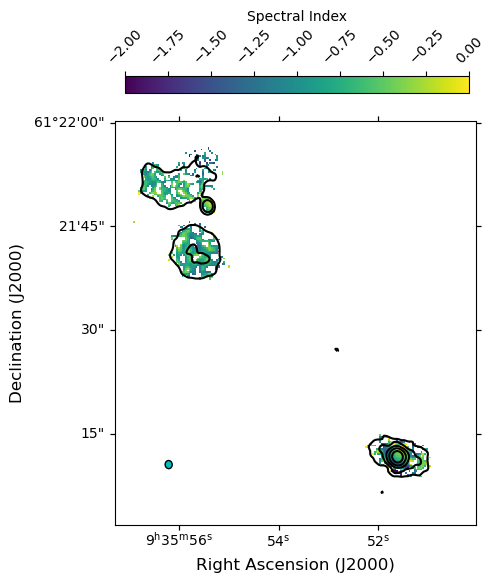}
    \caption{VLA L-band continuum (top) and spectral index (bottom) images of IRAS\,09320, with the nucleus located in the lower right. The upper left structure is a neighbor radio galaxy. The continuum is shown as grayscale-filled contours and the contour levels are (black) $62\ \mu\mbox{Jy}\ (5\sigma) \times (\pm 1, 2, 4, \ldots, 128)$ and (white) $62\ \mu\mbox{Jy}\ \times (256, 512, 1024)$. The peak surface brightness is 105.1~mJy~beam$^{-1}$. The spectral index image displays continuum contours at the same levels as in the top panel, with every other contour omitted to avoid obscuring the image. The inset in the top panel displays the GMOS H$\alpha+[$N\,{\sc ii}$]\lambda\lambda6548,84$ pseudo narrow-band image, with the radio contours overlaid in gray. The cyan-filled ellipse at lower left in both images represents the restoring beam which is of size $1.18''\times1.02''$ at a PA = 4$\degr$.}
    \label{fig:radio}
\end{figure}

\subsection{The VLA image} \label{32}
Fig.~\ref{fig:radio} shows, in the top panel, the VLA L-band continuum image of IRAS\,09320. A dominant radio core along with two-sided radio emission in the north-east-south-west direction at a position angle of $\sim70\degr$ in the sky is detected in IRAS\,09320. The total extent of the radio emission is $\sim10.5''$ or $\sim9$~kpc. An interesting-looking radio galaxy nearly $45''$ away to the north-east is also detected in the image. The peak surface brightness of the radio core in IRAS\,09320 is 105.1~mJy~beam$^{-1}$, while the total flux density of the entire source is 141.4~mJy. The bottom panel of Fig.~\ref{fig:radio} presents the in-band spectral index image of IRAS\,09320. The spectral index of the radio core is $-0.62\pm0.004$, consistent with optically thin emission from an AGN \citep[e.g.,][]{Kharb2006}. The brightness temperature of the kpc-scale radio core from the VLA image is $6.5\times10^5$~K, typical of an AGN \citep[e.g.,][]{Berton2018}. The average spectral index of the entire radio emission from IRAS\,09320 (including the core) is $-0.77\pm0.14$. It is noteworthy that the morphology as well as the spectral index of IRAS\,09320 is similar to that observed in radio-quiet Palomar Green (PG) quasars \citep[see][]{Silpa2023}.

\subsection{Gas kinematics}
In the second column of Fig.~\ref{fig:nii} we present the velocity fields for [N\,{\sc ii}]$\lambda$6583, once again with the first row corresponding to the narrow component and the second row to the broad component. The velocity fields are in units of km\,s$^{-1}$ and the systemic velocity of the galaxy, of 11\,772\,km\,s$^{-1}$ has been subtracted. A rotation pattern is clearly observed in the velocity map for the narrow component, with redshifts to the west of the nucleus and blueshifts to the east, and a projected velocity amplitude of 200\,km\,s$^{-1}$. Conversely, for the broad component, the velocity field exhibits blueshifts throughout, with negative velocities of $\approx-500$\,km\,s$^{-1}$. A slight gradient can be observed in the broad component, highlighted by the gray contours, which can be interpreted as resulting from an interaction between the outflow and the gas in galaxy's disk. Nevertheless, the outflow gas remains entirely blueshifted. We observe only the blueshifted portion of the outflow because its front side is visible; the absence of redshifted emission, corresponding to the far side of the outflow, is likely due to extinction caused by the galaxy's disk.

Still in Fig. \ref{fig:nii}, now in the third column, we have two-dimensional maps of the velocity dispersion for [N\,{\sc ii}]$\lambda6583$ emission line, in units of km\,s$^{-1}$. The narrow component exhibits values below 200\,km\,s$^{-1}$ throughout the field, with the highest values being observed at the nucleus and the lowest values, around $\sim100$\,km\,s $^{-1}$, seen at distances greater than 1\,arcsec from the nucleus, mainly in regions co-spatial with the spiral arm seen in the flux maps for the narrow component, reported above. The increase in velocity dispersion near the center could be partially attributed to beam smearing. However, it also aligns with the expected pattern of gas motion influenced by the galaxy's gravitational potential, characterized by higher velocity dispersion at the nucleus that decreases with distance from the center. The broad component presents velocity dispersion values ranging between $\sim500$ and $650$\,km\,s$^{-1}$. 

Thus, considering the observed gas kinematics and emission structure, we can clearly associate the narrow component to emission of gas located in the plane of the disk, while the broad component is produced by an ionized gas outflow.

\subsection{Emission-line ratio maps and diagnostic diagrams}

\begin{figure*}
    \includegraphics[width=\textwidth]{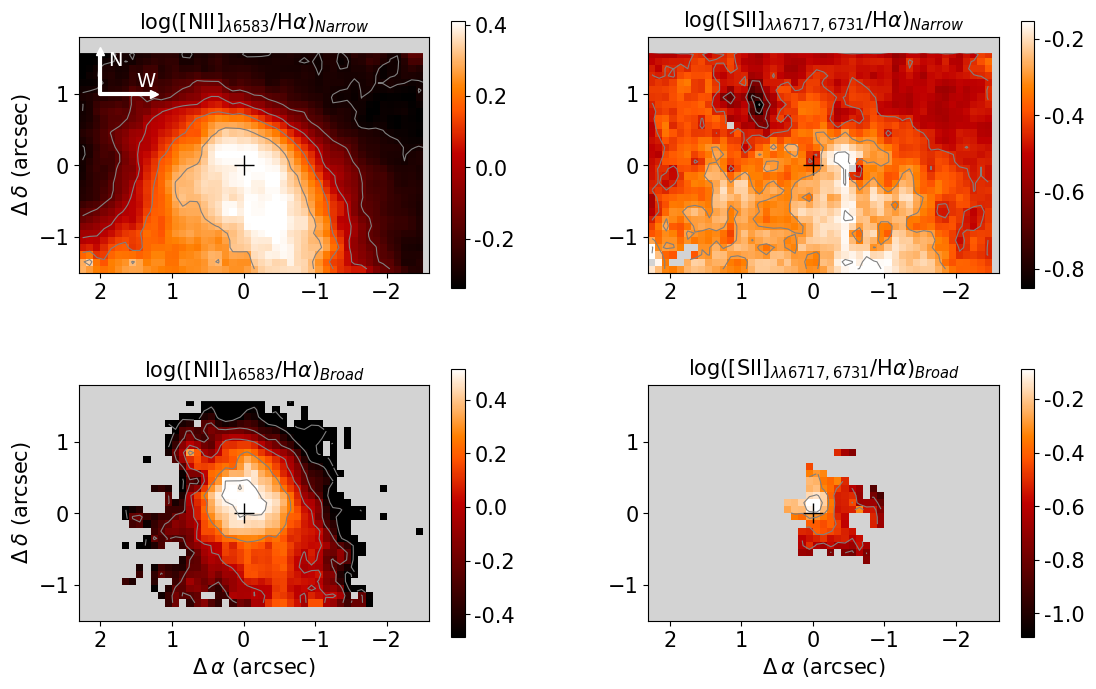}
    \caption{Ratio maps of emission line intensities for the narrow (top) and broad (bottom) components. On the left, in logarithmic units, ratios of [N\,{\sc ii}]$\lambda6583/$H$\alpha$ are shown, and on the right, ratios of [S\,{\sc ii}]$\lambda\lambda6717,6731/$H$\alpha$.}
    \label{fig:ratios}
\end{figure*}

Ratio maps of emission line intensity for the narrow and broad components are shown in Fig.~\ref{fig:ratios}. In the left column, we have maps of [N\,{\sc ii}]$\lambda6583/$H$\alpha$, and in the right column, we have maps of [S\,{\sc ii}]$\lambda\lambda6717,6731/$H$\alpha$. For the narrow component, the ranges of these ratios are $-0.1\lesssim\log$\:[N\,{\sc ii}]$\,\lambda6583$/H$\alpha\lesssim0.4$ and $-0.8\lesssim\log$\:[S\,{\sc ii}]$\,\lambda\lambda6717,6731$/H$\alpha\lesssim-0.2$. We observe that, for this component, both ratio maps have the highest values in the nuclear region, extending to the southwest, with an elongated morphology resembling a cone, typical of ionized gas outflows in AGNs. The lowest values for both line ratios are seen beyond the inner $\sim1.7$\,arcsec ($\sim1.4$\,kpc) in a region farther away from the nucleus, extending in a clockwise semicircular arc from east to southwest, close to the border of the FoV and encompassing the region where we detected a spiral arm of the galaxy and consistent with the values expected for emission of gas ionized by young stars. For the broad component, the [N\,{\sc ii}]$\,\lambda6583$/H$\alpha$ ratio map presents values of $-0.4\lesssim\log$\:[N\,{\sc ii}]$\,\lambda6583$/H$\alpha\lesssim0.5$, while the [S\,{\sc ii}]$\,\lambda\lambda6717,6731$/H$\alpha$ ratio map shows values of $-1.1\lesssim\log$\:[S\,{\sc ii}]$\,\lambda\lambda6717,6731$/H$\alpha\lesssim-0.3$.

\begin{figure*}
    \includegraphics[width=\textwidth]{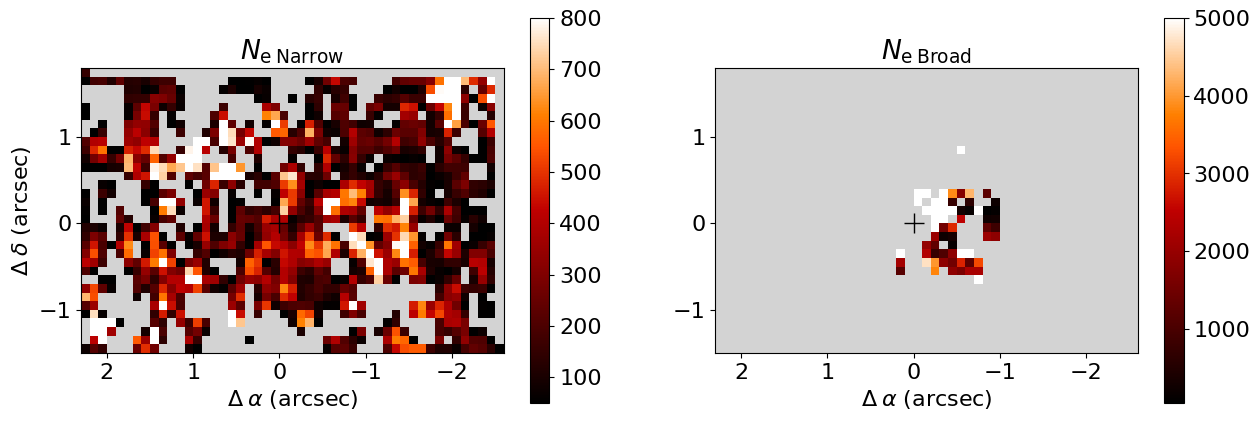}
    \caption{Electron density maps for the narrow (left) and the broad components (right), in units of number of electrons per cm$^3$. Gray regions correspond to locations where one or both emission lines were not detected above $3\sigma$ noise level in the continuum and locations where the [S\,{\sc ii}]$\lambda6716/6731$ values are outside the limits within which [S\,{\sc ii}] can be used to estimate N$_e$.}
    \label{fig:ne}
\end{figure*}

In Fig.~\ref{fig:ne} we present maps of the electron density ($N_e$), derived from the [S\,{\sc ii}]$\lambda6716/6731$ emission line ratio, by using the \textsc{PyNeb} code \citep{pyneb} and assuming an electron temperature of 15\,000\,K, a typical value observed for active galaxies \citep{riffel21a,dors17}. The gray regions in these maps correspond to locations where one or both emission lines were not detected above the 3$\sigma$ noise level in the continuum and locations where the [S\,{\sc ii}]$\lambda6716/6731$ values are outside the limits within which [S\,{\sc ii}] can be used to estimate $N_e$. The median density value for the disk component is $\sim215$\,cm$^{-3}$, while for the broad component the median density is $\sim2160$\,cm$^{-3}$. The electron density for the broad component, associated with the outflow, is approximately ten times higher than that of the narrow component, which corresponds to the gas in the galaxy's disk. This difference in $N_e$, with higher values in the outflow than in the disk, is expected and consistent with the observations \citep{Davies20,Revalski22,Binette24}, owing to the extreme conditions created by AGN-driven winds and jets, which enhance the density in the nucleus relative to the disk. In fact, the $N_e$ values could be even higher, as the [S\,{\sc ii}] lines used for the calculations arise from a partially ionized region, while the outflow is associated with fully ionized gas.

Emission-line ratio diagnostic diagrams are widely used tools to determine the mechanism of ionization and excitation of gas in galaxies and, thus, classify them. In this work, we use the BPT and WHAN diagrams \citep{bpt,fernandes11}.

The BPT diagram is a diagnostic diagram that allows us to investigate the ionization mechanism in galaxies based on emission line ratios, such as [O\,{\sc iii}]\,$\lambda$5007/H$\beta$ versus [N\,{\sc ii}]$\lambda6584$/H$\alpha$, or even [O\,{\sc iii}]\,$\lambda$5007/H$\beta$ versus [S\,{\sc ii}]$\lambda\lambda6717,6731$/H$\alpha$ diagrams. This diagram is divided into four distinct regions, as defined by \citet{kewley01}, \citet{kauffmann03} and \citet{scha07}, representing the different possible ionization mechanisms: strong AGN (e.g., Seyfert nuclei), weak AGN (e.g., LINERs), star-forming regions (e.g., Starbursts), and a transition or composite region (TO), which includes contributions from both AGN and star formation. As previously mentioned, we were able to detect the [O\,{\sc iii}]$\lambda5007$ and H$\beta$ emission lines only in a few spaxels around the nucleus. As a result, we were unable to construct a spatially resolved BPT diagram. By integrating the spectra within a circular aperture of 0.5\,arcsec in radius, centered on the nucleus, the resulting line ratios for [N\,{\sc ii}]$\lambda6584$/H$\alpha$ and [S\,{\sc ii}]$\lambda\lambda6717,6731$/H$\alpha$ are consistent with the presence of a Seyfert nucleus in IRAS\,09320+6134. This is consistent with the results obtained from higher S/N long-slit spectra (Wills et al., in prep.).

\begin{figure*}
    \includegraphics[width=\textwidth]{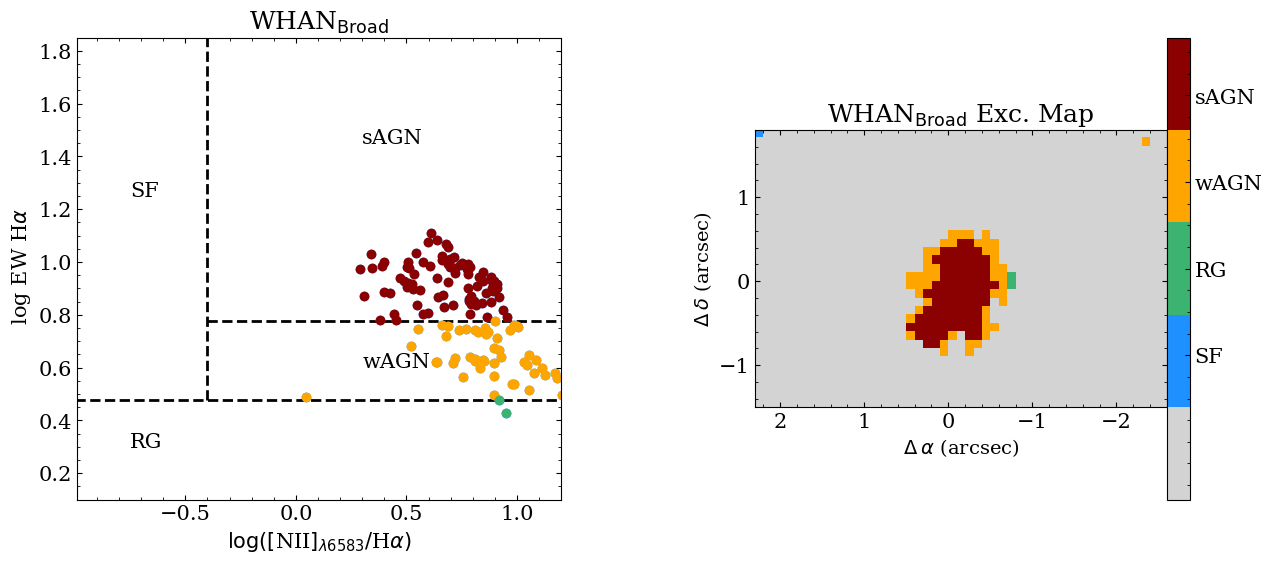}
    \caption{$\log$(EW\,H$\alpha$) versus $\log$([N\,{\sc ii}]$\lambda6583$/H$\alpha$) WHAN diagram for the broad component (left), corresponding to the ionized gas outflow, and respective excitation map (right). The dots represent the values for each spaxel and the lines represent the separation limits between the different gas ionization mechanisms.}
    \label{fig:whan}
\end{figure*}

Fig. \ref{fig:whan} presents the WHAN diagram, which, similarly to the BPT diagram, also allows analysis and a distinction to be made as to whether the ionization mechanism is due to an AGN or Starburst. It is particularly useful to separate regions where the gas excitation is due to a true AGN from regions where the gas emission is produced by hot low-mass evolved stars (HOLMES), i.e. fake AGN. An advantage of the WHAN diagram is that only the H$\alpha$ and [N\,{\sc ii}]$\lambda6584$ emission lines are necessary. The WHAN diagram for the narrow component shows that all spaxels are located in the strong AGN region and thus, we do not show it here. This, together with the result obtained for the BPT diagram, indicate that the main ionizing source of the gas in the disk is an AGN. The WHAN diagram for the broad component is shown in Fig. \ref{fig:whan}, along with its excitation map. Again, most of the points are observed in the AGN region of the diagram, with the central locations of the galaxy being in the strong AGN region of the diagram, surrounded by points in the weak AGN region. This indicates that the ionization of the gas emission in outflow is also due to an AGN. Therefore, combining the results from the BPT and WHAN diagrams, we can conclude that the main ionization of both kinematic components, disk and outflow, is due to an AGN.

\section{Discussion} \label{4}

As presented above, the gas kinematics in the central 3.0\:$\times$\:4.2\:kpc$^2$ of IRAS\:09320 clearly present two components, one due to the emission of gas in the plane of the disk and another due to an AGN driven outflow. The main excitation mechanism of both components seems to be photoionization by the AGN radiation field, as indicated by the emission-line ratio maps and the BPT and WHAN diagnostic diagrams. In addition, the emission-line flux distributions reveal a spiral arm of star-forming regions, particularly observed in H$\alpha$. In this section, we characterize and discuss the disk, outflow and star-forming regions properties and put the results found for IRAS\:09320 in context with those from previous studies.

\subsection{The disk component}

In order to describe and characterize the gas disk in IRAS\,09320, we use the {\sc kinemetry} method \citep{kinemetry} applied to the narrow component of the [N\,{\sc ii}]$\lambda6583$ velocity field. This method proceeds on the assumption that it is possible to characterize the observed velocity field assuming that the gas is distributed along a set of ellipses, with the gas distribution along the ellipse satisfying a cosine law, while the velocity field itself is modeled as circular motion within a thin disk. Applying {\sc kinemetry} to the velocity field, it provides the modeled velocity field along with the ellipse’s parameters, such as the position angle ($\Psi_0$), ellipticity, and kinemetric coefficients k$_n$ as a function of radius.

\begin{figure*}
    \includegraphics[width=\textwidth]{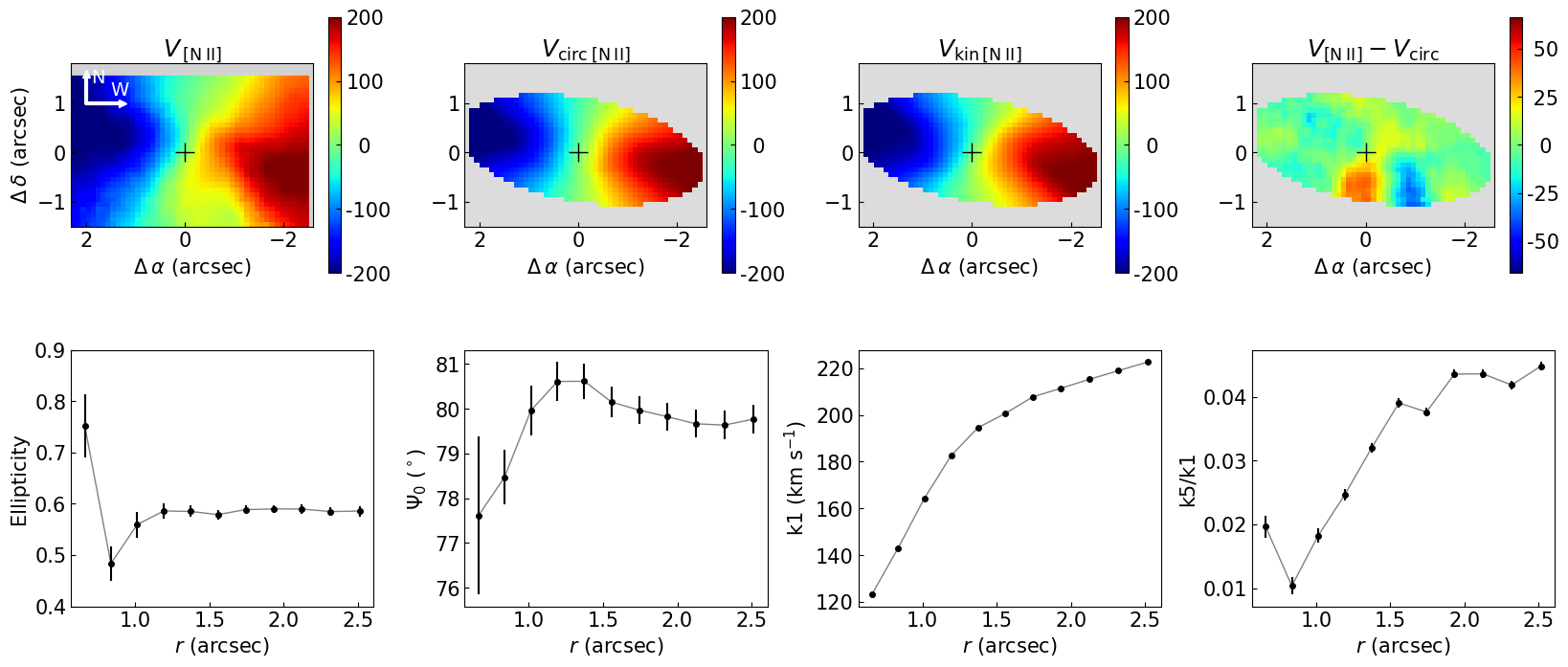}
    \caption{[N\,{\sc ii}] velocity maps (top) and kinemetric coefficients (bottom) resulting from the {\sc kinemetry} fit. On the top panels, from left to right, we present maps of the observed gas velocity field for the narrow component, circular velocity map, kinemetric best-fitting velocity model and the residual velocity, which is obtained by subtracting the observed gas velocity and the fitted circular velocity. On the bottom panel, from left to right, are the ellipticity, semi-major axis position angle and the harmonic expansion coefficients k$_1$ and k$_5$/k$_1$.}
    \label{fig:kinemetry}
\end{figure*}

Figure~\ref{fig:kinemetry} shows [N\,{\sc ii}] velocity maps and kinemetric parameters resulting from the {\sc kinemetry} fit. In the top panels, from left to right, are the observed [N\,{\sc ii}] velocity field for the narrow component, associated with the gas emission in the disk, the circular velocity map, kinemetric best-fitting velocity model and the residual velocity, obtained by subtracting the model from the observed velocities. The residuals are smaller than 20\:km\:s$^{-1}$ at most locations. Some residuals larger than 20\:km\:s$^{-1}$ are seen only to the south of the nucleus, a region where the broad component emission is more extended (Fig.~\ref{fig:nii}) and the emission line ratio maps present higher values (Fig.~\ref{fig:ratios}). Therefore, we interpret these residuals as resulting from turbulence caused by the interaction of the outflow with the gas in the disk. This is further supported by the higher velocity dispersion observed in the narrow-line component within this region, as illustrated in Fig.~\ref{fig:nii}.

The bottom panels of Fig.~\ref{fig:kinemetry} present, from left to right, the ellipticity, the radial variation of the kinematic position angle of the major axis of the fitted ellipses, and the harmonic coefficients $k_1$ and $k_5/k_1$. The coefficient $k_1$ describes the velocity amplitude of bulk gas motions and the ratio $k_5/k_1$ provides the contribution of non-rotational motion relative to the rotational component \citep{kinemetry}.

The ellipticity shows a slight variation in the innermost region ($<1$\,arcsec), but remains constant at $\sim0.6$ for radii greater than 1\,arcsec. Similarly, the position angle does not exhibit significant variations, indicating that the galaxy's kinematic major axis is constant at all observed radii, with a value of $\sim80^{\circ}$. This is consistent with the value of $84.3^{\circ}$ determined for the large-scale disk in previous works \citep[e.g.][]{positionangle}. Both the $k_1$ and $k_5/k_1$ coefficients show increasing values from the nucleus to larger radii. Although the values for the $k_5/k_1$ ratio increase, the maximum value reached is still very low, of around 0.04. This ratio is used to identify non-circular or non-rotational movements, thus, the low values obtained for IRAS\,09320 indicate that the narrow component is dominated by the gas in rotation in the plane of the galaxy's disk.

Thus, we conclude that the origin of the narrow component in the emission lines is gas rotating in the plane of the galaxy's disk, but ionized by the radiation field of the AGN, as indicated by the emission-line ratio diagnostic diagrams for this component.

\subsection{Outflow properties}

From the emission-line flux distribution and kinematics of the broad component, we can estimate the outflow properties, i.g. the mass outflow rate and the kinetic power of the ionized outflow. The recipe to determine the mass outflow rate, i.e., the ratio between the mass of ionized gas in outflow and the dynamic time for the gas to reach its current position is given in \cite{heka2024} and is described as follows. The mass of gas can be estimated by
\begin{equation}
    M_{\rm o}=N_{\rm e}m_{\rm p}V f,
\end{equation}
where $N_{\rm e}$ is the electron density, $m_{\rm p}$ is the mass of the proton, $V$ is the volume of the region occupied by the outflow and $f$ is the filling factor, which can be estimated from the H$\alpha$ luminosity emitted within the volume $V$ using the equation
\begin{equation}
    L_{\rm H\alpha}\approx f N_{\rm e}^2j_{\rm H\alpha}(T)V,
\end{equation}
where $j_{\rm H\alpha}$ is the emission coefficient given by $j_{\rm H\alpha}(T)=3.3534\times10^{-25}$\,erg\,cm$^{-3}$\,s$^{-1}$ \citep{oster}. Combining these two equations, the mass of ionized gas in outflow can be obtained from
\begin{equation}
    M_{\rm o}=\dfrac{m_{\rm p}L_{\rm H\alpha}}{N_{\rm e}j_{\rm H\alpha}(T)}.
\end{equation}

The observed H$\alpha$ luminosity of the broad component is $L_{\rm H\alpha b}=(2.9\pm0.12)\times10^{39}$\,erg\,s$^{-1}$. The uncertainty in flux is obtained by assuming a Gaussian noise distribution in the continuum, where a rough estimate of the flux uncertainty is given by $\sigma_{\rm cont}\times$FWHM, where the $\sigma_{\rm cont}$ is the standard deviation measured in a region close to the emission line and FWHM is the line full width at half maximum. We can estimate the visual extinction ($A{\rm _V}$) using the H$\alpha$/H$\beta$ ratio for the broad component, following the recipe presented in \citet{rogerioriffel21}. As the H$\beta$ emission line is detected only in a few individual spaxels, we use the integrated spectra within a circular aperture of 0.5\,arcsec radius centred at the nucleus to estimate the emission lines fluxes, resulting in A$_V=4.5\pm0.8$\,mag for the outflowing gas. 
The resulting extinction corrected H$\alpha$ luminosity is $L_{\rm H\alpha b,corr}= (8.7\pm0.6)\times10^{40}$\,erg\,s$^{-1}$ by adopting the extinction law from \cite{cardelli89}. Using the median value of the electron density for the outflow component, $\langle N_{\rm e}\rangle=2160\pm300$\,cm$^{-3}$, and the extinction corrected H$\alpha$ luminosity, the resulting mass of ionized gas in the outflow is $M_{\rm o}= (1.01\pm0.21)\times10^5\,\text{M}_{\odot}$.

Now, to estimate the mass-outflow rate, we determine the dynamical time assuming a radius of $r=0.5$\,arcsec $\approx 424$\,pc, which corresponds to roughly half of the FWHM of the flux distribution of the [N\,{\sc ii}]$\lambda6583$ broad component, and the outflow velocity of $V_{\text{out}}\approx500$\,km\,s$^{-1}$, obtained directly from the velocity field of the [N\,{\sc ii}]$\lambda6583$ broad component (Fig. \ref{fig:nii}), and which represents a lower limit for the outflow velocity, as it is not corrected for projection effects due to the unknown geometry of the outflow. Finally, the mass outflow rate is
\begin{equation}
    \dot{M}_{\rm o}=\dfrac{M_{\rm o}V_{\text{out}}}{r}=0.122\pm0.026\text{\,M}_{\odot}\,\text{yr}^{-1}.
\end{equation}

The kinetic power of the outflow can be determined using
\begin{equation}
    \dot{E}_{\text{out}}=\dfrac{1}{2}\dot{M}_{\rm o}(V^2_{\text{out}}+3\sigma^2_{\text{out}}),
\end{equation}
where $\dot{M}_{\rm o}$ is the derived mass outflow rate, $V_{\text{out}}$ is the outflow velocity and $\sigma_{\text{out}}$ is its velocity dispersion, these last two derived from Fig. \ref{fig:nii}. Using $V_{\text{out}}\approx500$\,km\,s$^{-1}$ and $\sigma_{\text{out}}\approx600$\,km\,s$^{-1}$ the resulting kinetic power of the outflow is $\dot{E}_{\text{out}}=(5.1\pm0.1)\times10^{40}$\,erg\,s$^{-1}$. 

The kinetic power of the outflow can be compared to the AGN bolometric luminosity ($L_{\text{bol}}$), which can be estimated from the [O\,{\sc iii}]$\lambda$5007 luminosity, as $L_{\text{bol}}=3500$L$_{\text{[O\,{\sc iii}]}}$ \citep{heckman05}.
We determined $L_{\text{[O\,{\sc iii}]}}$ in the same way as was calculated H$\alpha$ luminosity above, and obtained the observed [O\,{\sc iii}] luminosity of the broad component of $L_{\text{[O\,{\sc iii}]}\text{,b}}\approx5.2\times10^{38}$\,erg\,s$^{-1}$. After the extinction correction, the resulting [O\,{\sc iii}] luminosity is L$_{\text{[O\,{\sc iii}]}}\approx5.4\times10^{40}$\,erg\,s$^{-1}$, so that $L_{\text{bol}}\approx1.8\times10^{44}$\,erg\,s$^{-1}$. Thus, the kinetic efficiency of the outflow is $\epsilon=\dot{E}_{\text{out}}/L_{\text{bol}}\approx2.8\times10^{-4}$.

The value found for the outflow's kinetic efficiency is much lower than the value required by cosmological simulations for the outflow to be efficient in suppressing star formation in the host galaxy and, therefore, affecting the galaxy's evolution. It is estimated that the kinetic power of the outflow must be at least 0.5\% of the AGN bolometric luminosity so that the outflow reaches kpc scales and is capable of efficiently cleaning the cold gas from the galaxy's nucleus \citep{hopkins10,harrison18}.

\begin{figure}
    \includegraphics[width=\linewidth]{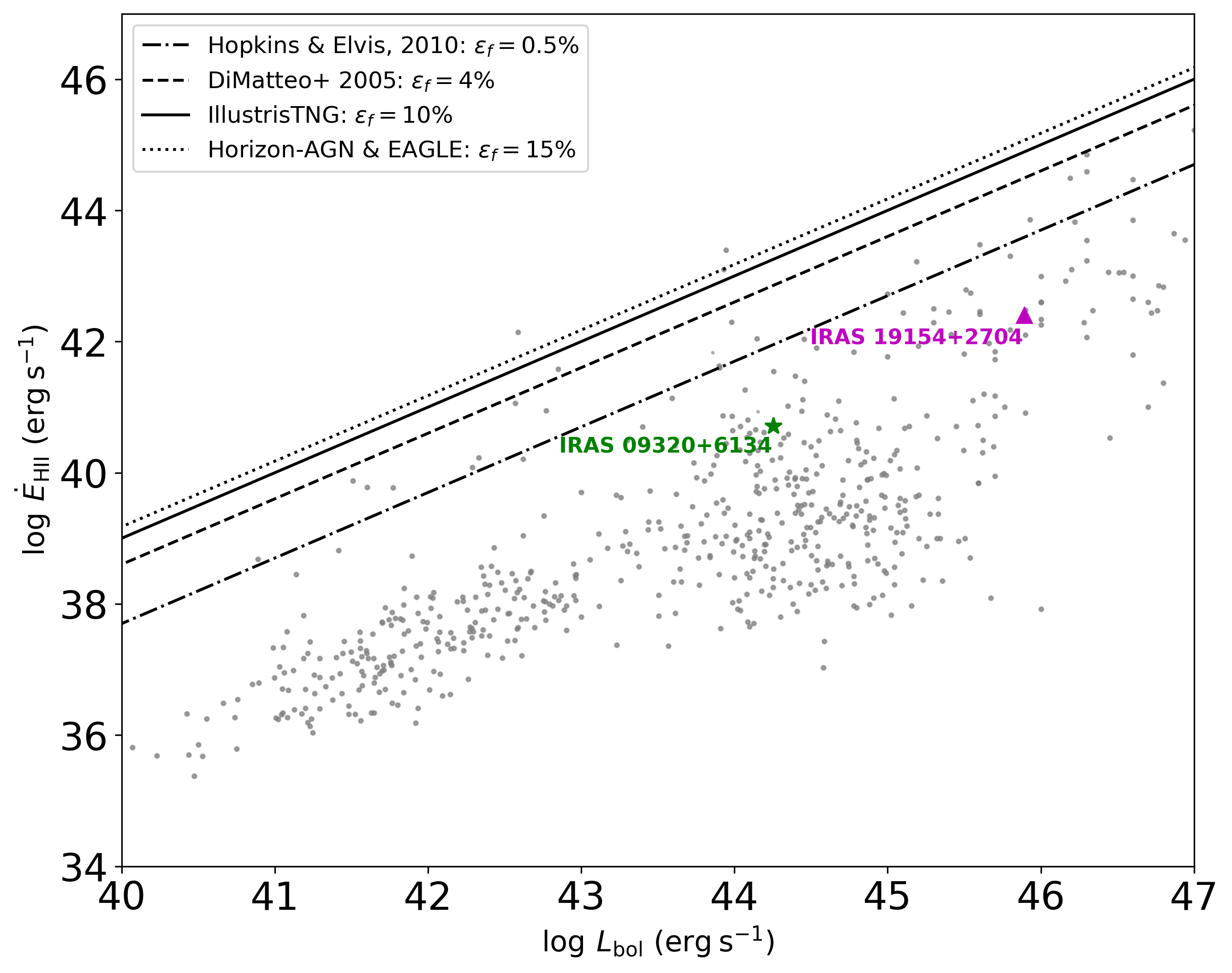}
    \caption{AGN bolometric luminosity versus kinetic power of outflows for a sample of over 600 galaxies compiled from the literature \citep{riffel21b,riffel24,hermosa24,ulivi24}, represented by small gray dots. The lines indicate the minimum outflow coupling efficiencies required to effectively suppress star formation, as predicted by different simulations \citep{dimatteo05,hopkins10,dubois14,harrison18,xu22}. IRAS\,09320 is indicated as a green star. IRAS\,19154 is also shown, as a lilac triangle, based on the results derived from \citet{heka2024}.}
    \label{fig:potvslbol}
\end{figure}

Figure~\ref{fig:potvslbol} presents a plot of AGN bolometric luminosity versus the kinetic power of ionized outflows, based on measurements of more than 600 galaxies and covering seven orders of magnitude in AGN luminosity. The dataset is primarily drawn from \citet{riffel21b}, updated to include more recent studies (e.g., \citealt{riffel24,hermosa24,ulivi24}). Lines indicate the minimum outflow coupling efficiencies required to effectively suppress star formation, as predicted by different simulations \citep{dimatteo05,hopkins10,dubois14,harrison18,xu22}. In addition to the galaxy IRAS\,09320, we also include IRAS\,19154+2704 in the figure, which was the sixth OHMG analyzed from our sample and the first in which a sufficiently strong outflow was detected, allowing us to investigate its properties \citep{heka2024}. The mass outflow rate and kinetic power of the ionized outflow in IRAS\,19154 are estimated to be $\dot{M}_{\text{out}}\approx(4\pm2.6)$\,M$_{\odot}$\,yr$^{-1}$ and $\dot{E}_{\text{out}}=(2.5\pm1.6)\times10^{42}$\,erg\,s$^{-1}$, respectively. Similar to IRAS\,09320, the outflow observed in IRAS\,19154 is also AGN-driven. For both IRAS\,09320 and IRAS\,19154, the observed kinetic coupling efficiencies fall below the minimum thresholds predicted by simulations.

Nonetheless, care must be taken when comparing observations with theoretical models, as the models suggest that most of the outflow power arises not from kinetic energy transfer, as calculated here, but rather from emitted radiation \citep{harrison18,Harrison24}. Outflows are also typically observed in multiple gas phases, meaning that the ionized gas component represents only a fraction of the total outflow. Moreover, recent theoretical studies suggest that low-power outflows can effectively suppress star formation in the host galaxy if sustained over timescales of about $\sim1$\,Myr \citep{almeida23}. If the outflow rate remains constant throughout an entire AGN duty cycle of 1 million years \citep{Novak11}, it could expel approximately 10$^5$ M$_\odot$ of gas from the central region of IRAS\,09320. Although this process effectively redistributes gas within the galaxy, it will still remain available for future star formation.

\subsection{Comparison with previous studies}

IRAS\:09320 is the seventh galaxy in our sample where we conducted a detailed study using multiwavelength observations. The first study of this series was performed by \citet{sales15} for the galaxy IRAS\,16399-0937, a merger in mid to late stage, with two nuclei immersed in a diffuse envelope. An AGN embedded in dust was detected in one of the nuclei while the other is dominated by starburst activity. Next, \cite{heka2018a} studied galaxy IRAS\,23199+0123, an interacting pair where a Seyfert 1 nucleus was detected in the western member. A previously unknown maser emission was also discovered and reported. A correlation between the maser sources and shocks generated by outflows from the AGN was identified. In the third study, \cite{heka2018b} presented IRAS\,03056+2034, a barred spiral galaxy with irregular structures and flocculent spiral arms, indicative of an interaction that occurred in the past. A central AGN immersed in dust and surrounded by a ring of regions of star formation have been detected. Continuing, IRAS\,17526+3253 was investigated in \cite{sales19}. This object consists in a significant mid-stage merger, hosting OH and H$_2$O masers. The gas ionization was attributed to star formation regions, yet no conclusive determination was made regarding the existence of an AGN. However, this possibility cannot be dismissed based only on the analysis of the obtained data. In fifth study, \cite{heka2020} presented IRAS\,11506-3851, an isolated spiral galaxy with a two-bar structure and no clear signs of interaction. It was detected the presence of a weak AGN embedded in a region dominated by star formation around the nucleus, along with the presence of a modest outflow originating from the AGN. In the last and most recent study, \cite{heka2024} investigated IRAS\,19154+2704, an irregular galaxy with clear signs of interaction, but apparently with only one nucleus, indicating that the interaction is in an advanced stage and was likely the trigger of the AGN detected in the galaxy. An outflow of ionized gas was also detected in this galaxy, the first one that was possible to study more deeply as it was the most intense one detected among the galaxies studied up to that point. This galaxy is the only one in which the OH maser appears in absorption, while the previous ones are all classified as OHM sources.

Finally, in this work, we examine the distribution, kinematics, and excitation of the gas in the galaxy IRAS\,09320+6134. This galaxy exhibits signs of past interactions and clearly shows the presence of an AGN, as well as circumnuclear star-forming regions, such as northwest of the nucleus. Photoionization by the AGN radiation field explains both the emission of gas in the galaxy's disk and in the outflow.

Combining the results of this study with those of the other six galaxies in the sample, we find that, in terms of structure, all galaxies exhibit signatures of interaction. Three of them are in the process of merging, and four appear to be the result of a merger that occurred in the recent past. Regarding the ionization mechanism of the gas, the presence of an AGN was detected in six of the seven galaxies, and there seems to be a pattern of star-forming regions around the nucleus, on scales of a few kiloparsecs, in most of the objects. Outflows in ionized gas were detected in four out of seven galaxies. In all cases, the kinetic coupling efficiency of the outflows is low, similar to the case of IRAS\,09320+6134, below the value estimated by cosmological simulations for the outflow to significantly impact the environment and evolution of their host galaxies. 
However, it is improbable that all the energy injected by the AGN is in the form of the kinetic power within the outflow. Simulations suggest that the kinetic energy of the outflows constitutes less than 20\% of the total emitted outflow energy \citep{richings18b}. Thus, even though the power of the outflows we have been finding are low, when combined with the associated feedback from the AGN radiation, will end up affecting the star formation in the host galaxy \citep{almeida23}.

\section{Conclusions} \label{5}
We conducted a two-dimensional mapping of the gas distribution, kinematics, and excitation in the central 3\,kpc $\times$ 4.2,kpc region of the OH megamaser galaxy IRAS\,09320, using GMOS-IFU observations. The GMOS data are complemented by larger scale HST and VLA observations. Our main conclusions are as follows:

\begin{enumerate}
    \item HST ACS F814W - i band and H$\alpha+[$N\,{\sc ii}$]\lambda\lambda6548,84$ narrow-band images suggest that IRAS\,09320 is a late-stage merger by the detection of a single bright nucleus;
    \item The VLA radio image reveals a dominant radio core along with two-sided radio emission in the NE-SW direction. The spectral index and the brightness temperature of the radio core are consistent with emission from an AGN. Also, the extended emission seen in IRAS\,09320 resembles in morphology and spectral index the emission observed in radio-quiet PG quasars.
    \item GMOS data reveal that the ionized gas emission in the inner $\sim2$\,kpc radius presents two kinematic components: a narrow component, with velocity dispersions of $100\leq\sigma\leq200$\,km\,s$^{-1}$, and a broad component, with velocity dispersions of $500\leq\sigma\leq650$\,km\,s$^{-1}$;
    \item The narrow component traces gas emission located in the disk plane and is consistent with a rotating disk, exhibiting velocities of approximately $\sim200$\:km\:s$^{-1}$, aligned with the orientation of the galaxy's large-scale disk;
    \item The broad component is attributed to an AGN-driven outflow, with bulk velocities of up to $500$\,km\,s$^{-1}$, observed only in blueshifts;
    \item Emission-line ratio BPT and WHAN diagrams indicate gas excitation by a Seyfert-type AGN, which is probably the driver of the outflow. Ratio maps of emission line intensity, along with flux maps, also indicate the presence of star-forming regions associated with spiral arms, to the northwest and southeast of the nucleus;
    \item The mass of ionized gas in outflow is $M_{\rm o}= (1.01\pm0.21)\times10^5\,\text{M}_{\odot}$, corresponding to a mass outflow rate of $\dot{M}_{\rm o}=0.122\pm0.026\text{\,M}_{\odot}\,\text{yr}^{-1}$. The kinetic coupling efficiency of the outflow, $\epsilon\approx2.8\times10^{-4}$, is below the value estimated by cosmological simulations for the outflow to impact the environment and evolution of their host galaxy. However, the outflow may be able to redistribute the gas within the inner region of the galaxy, adding to the feedback from the AGN radiation.
\end{enumerate}

Our study of the gas kinematics and excitation in galaxies with OH megamaser emission, concerning the galaxy IRAS\,09320+6134 and the other galaxies studied so far, has shown a connection between these objects and the presence of AGN activity, contrary to what was previously thought, that galaxies with OH megamaser are mainly associated with starburst activity. What we have observed is that in galaxies where the merger stage is not so advanced, indeed starbursts dominate the ionization, often found in circumnuclear regions. However, for galaxies in more advanced merger stages, as is the case with IRAS\,09320+6134, AGN activity proves to be dominant in the gas ionization mechanism in the galaxy, in some cases even strong enough to cause outflows of ionized gas. Combined with the radiative feedback from the AGN, the outflows may disrupt or delay star formation processes within the inner regions (few kpc) of the galaxy.

\section*{Acknowledgements}

We thank an anonymous referee for their comments and suggestions, which helped us to improve the present paper. 
CC thanks the financial support from Coordena\c c\~ao de Aperfei\c coamento de Pessoal de N\'ivel Superior - Brasil (CAPES) - Finance Code 001.
RAR acknowledges the support from Conselho Nacional de Desenvolvimento Cient\'ifico e Tecnol\'ogico (CNPq; Proj. 303450/2022-3, 403398/2023-1, \& 441722/2023-7), Funda\c c\~ao de Amparo \`a pesquisa do Estado do Rio Grande do Sul (FAPERGS; Proj. 21/2551-0002018-0), and CAPES (Proj. 88887.894973/2023-00).
DAS acknowledges CNPq and FAPERGS.
CO and SB acknowledge support from the Natural Sciences and Engineering Research Council (NSERC) of Canada.

Based on observations obtained at the Gemini Observatory, which is operated by the Association of Universities for Research in Astronomy, Inc., under a cooperative agreement with the NSF on behalf of the Gemini partnership: the National Science Foundation (United States), National Research Council (Canada), CONICYT (Chile), Ministerio de Ciencia, Tecnolog\'{i}a e Innovaci\'{o}n Productiva (Argentina), Minist\'{e}rio da Ci\^{e}ncia, Tecnologia e Inova\c{c}\~{a}o (Brazil), and Korea Astronomy and Space Science Institute (Republic of Korea). This research has made use of NASA's Astrophysics Data System Bibliographic Services. This research has made use of the NASA/IPAC Extragalactic Database (NED), which is operated by the Jet Propulsion Laboratory, California Institute of Technology, under contract with the National Aeronautics and Space Administration.

\section*{DATA AVAILABILITY}

The GEMINI data used in this work is publicly available online via the GEMINI archive \url{https://archive.gemini.edu/searchform/}, with project code GN-2021B-Q-319. The HST data is available at \url{https://archive.stsci.edu/hst/}, with project code 11604.
The VLA data is available at \url{https://science.nrao.edu/facilities/vla/archive}, with project code 16B-063.
The data cubes and maps produced from these data can be shared on reasonable request to the corresponding author.





\bibliographystyle{mnras}
\bibliography{refs}





\appendix

\section{Gas distribution and kinematics for the other emission lines}

\begin{figure*}
    \includegraphics[width=15.3cm]{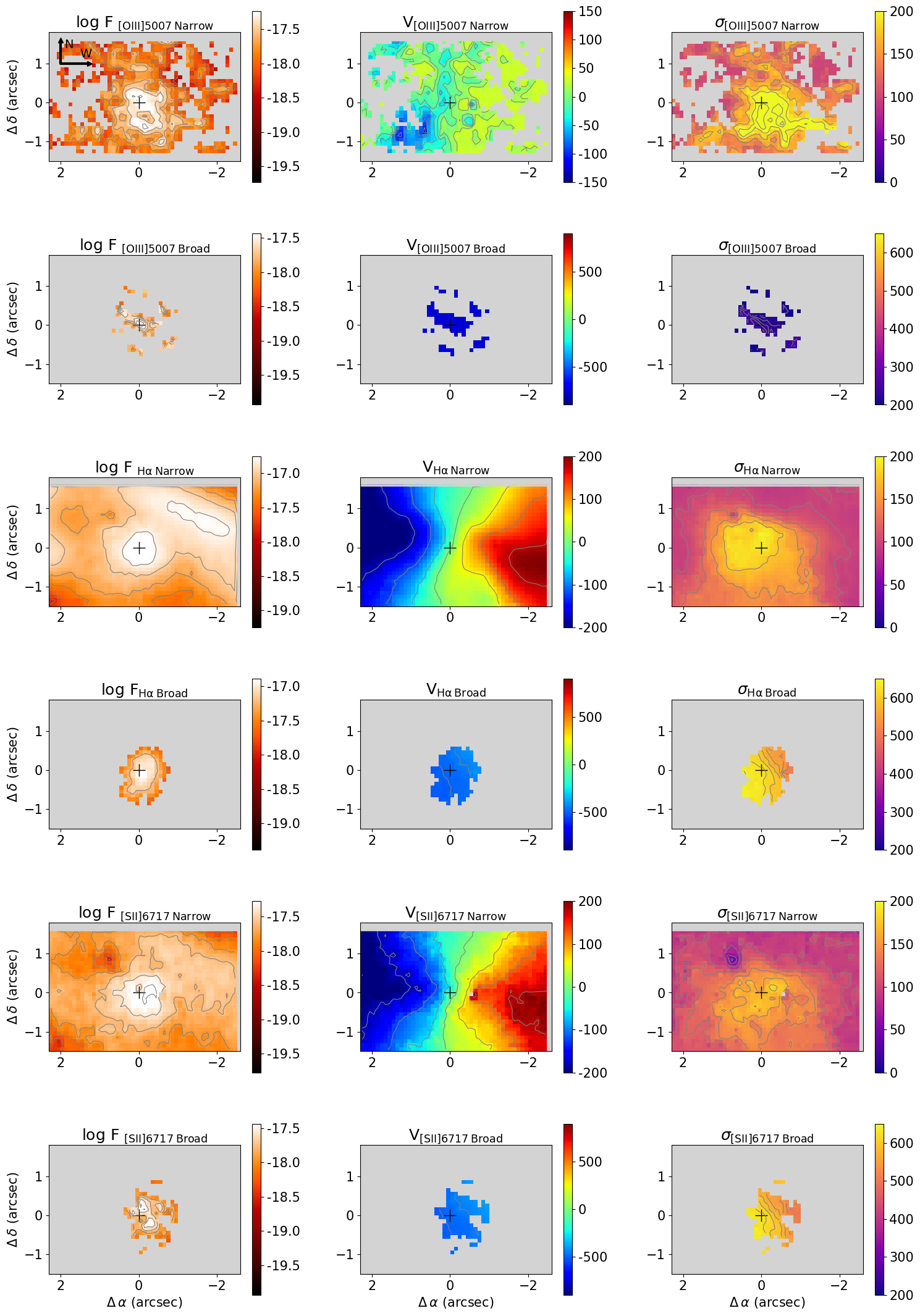}
    \caption{Maps of the distribution and kinematics of the gas in the galaxy IRAS\:09320, for the emission lines [O\,{\sc iii}]$\lambda5007$, H$\alpha$ and [S\,{\sc ii}]$\lambda$6717. Odd rows show the maps for the narrow component, while even rows correspond to the broad component. Flux distributions are shown in the first column, in logarithmic units of erg\,s$^{-1}$\,cm$^{-2}$\,spaxel$^{-1}$. In the second and third columns are the maps of velocity fields and velocity dispersion, respectively, in units of km\,s$^{-1}$. Velocity fields are shown after subtraction of the galaxy's systemic velocity. In all maps, the central cross marks the position of the galaxy's nucleus, and gray areas correspond to masked regions where the emission line was not detected or the signal-to-noise ratio was not high enough to allow measurements.}
    \label{fig:hao3s2}
\end{figure*}

Figure \ref{fig:hao3s2} shows the flux distributions, velocity fields and velocity dispersions for the narrow (odd rows) and broad (even rows) components of the [O\,{\sc iii}]$\lambda5007$, H$\alpha$ and [S\,{\sc ii}]$\lambda$6717 emission lines, respectively, similar to the Figure \ref{fig:nii}. In the first column, corresponding to the flux maps, all emission lines show their peak emission at the nucleus, for both components. Also, the northwestern region with high flux values in the narrow H$\alpha$ map corresponds to the region we associate with a possible spiral arm of the galaxy. The broad component is confined to a region of just under 1\,arcsec from the nucleus for all emission lines, while the narrow component shows extended emission throughout the field, particularly for the H$\alpha$ and [S\,{\sc ii}]$\lambda$6717 lines, as well as for the [O\,{\sc iii}]$\lambda5007$ line, although the map for this emission line is noisier compared to the others, due to its detection in a smaller number of spaxels. In the second column, the velocity fields of H$\alpha$ and [S\,{\sc ii}]$\lambda6717$ are very similar to those of [N\,{\sc ii}]$\lambda6583$, exhibiting a rotation pattern for the narrow component with amplitudes of 200\,km\,s$^{-1}$, and blueshifts for the broad component, with negative velocities of $\approx-500$\,km\,s$^{-1}$. For the [O\,{\sc iii}]$\lambda5007$ line, the velocity field of the narrow component also suggests a rotational pattern, but with lower amplitudes, around 50\,km\,s$^{-1}$. Meanwhile, the broad component also exhibits blueshifts, with negative velocities of $\approx-800$\,km\,s$^{-1}$. Finally, the third column displays the velocity dispersion maps. The narrow component shows values below 200\,km\,s$^{-1}$ throughout the field, with the highest values observed at the nucleus and the lowest, around 50\,km\,s$^{-1}$, at distances greater than 1\,arcsec from the nucleus, particularly in regions co-spatial with the structure attributed to a spiral arm of the galaxy. The broad component exhibits higher values, ranging from 500 to 650\,km\,s$^{-1}$ for the H$\alpha$ and [S\,{\sc ii}]$\lambda6717$ lines, and around 250\,km\,s$^{-1}$ for the [O\,{\sc iii}]$\lambda5007$ line.

\bsp	
\label{lastpage}
\end{document}